\documentclass[fleqn,usenatbib]{mnras}

\usepackage{newtxtext,newtxmath}
\usepackage[T1]{fontenc}

\DeclareRobustCommand{\VAN}[3]{#2}
\let\VANthebibliography\thebibliography
\def\thebibliography{\DeclareRobustCommand{\VAN}[3]{##3}\VANthebibliography}

\usepackage{graphicx}	 % Including figure files
\usepackage{amsmath}	 % Advanced maths commands
\usepackage{academicons} % Orcid handling
\newcommand{\gaia}{\textit{Gaia}}
\newcommand{\kepler}{\textit{Kepler}}
\newcommand{\ktwo}{\textit{K2}}
\newcommand{\tess}{\textit{TESS}}
\newcommand{\logg}{log~$g$}
\newcommand{\teff}{T$_{\rm{eff}}$}
\newcommand{\dnu}{$\Delta\nu$}

\newcommand{\cannon}{\texttt{The Cannon}}
\def\orcid#1{\kern .08em\href{https://orcid.org/#1}{\includegraphics[keepaspectratio,width=0.7em]{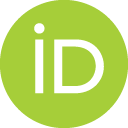}}}

\title[Key Spectroscopic Indicators of Red Giant Evolution]{CN and CO Features: Key Indicators of Red Giant Evolutionary Phase in Moderate-Resolution X-Shooter Spectra}
\author[K. A. Banks et al.]{Kirsten A. Banks$^{1,2}$\thanks{E-mail: k.banks@unsw.edu.au} \orcid{0000-0001-5210-1696},
Chantel Y. Y. Ho$^{1}$ ,
Sarah L. Martell$^{1,2,3}$ \orcid{0000-0002-3430-4163},
Sven Buder$^{2,4}$ \orcid{0000-0002-4031-8553},
Dennis Stello$^{1,2}$ \orcid{0000-0002-4879-3519},\newauthor
Sanjib Sharma$^{5}$ \orcid{0000-0002-0920-809X},
James Priest$^{1}$ \orcid{0000-0003-2380-2623},
Ana\"{i}s Gonneau$^{6}$ \orcid{0000-0001-9091-5666},
Keith Hawkins$^{7}$ \orcid{0000-0002-1423-2174}
\\
$^{1}$School of Physics, University of New South Wales, Sydney, NSW 2052, Australia\\
$^{2}$Centre of Excellence for All-Sky Astrophysics in Three Dimensions (ASTRO 3D), Australia\\
$^{3}$UNSW Data Science Hub (uDASH), University of New South Wales, Sydney, NSW 2052, Australia\\
$^{4}$Research School of Astronomy and Astrophysics, Australian National University, Canberra, ACT 2611, Australia\\
$^{5}$ Space Telescope Science Institute, Baltimore, MD 21218, USA\\
$^{6}$Institute of Astronomy, University of Cambridge, Madingley Road, Cambridge CB3 0HA, UK\\
$^{7}$Department of Astronomy, The University of Texas at Austin, 2515 Speedway Boulevard, Austin, TX 78712, USA
}

\date{Sumbitted to MNRAS May 2023}

\pubyear{2023}

\begin{document}
\label{firstpage}
\pagerange{\pageref{firstpage}--\pageref{lastpage}}
\maketitle

\begin{abstract}
Data-driven analysis methods can help to infer physical properties of red giant stars where ``gold-standard'' asteroseismic data are not available. The study of optical and infrared spectra of red giant stars with data-driven analyses has revealed that differences in oscillation frequencies and their separations are imprinted in said spectra. This makes it possible to confidently differentiate core-helium burning red clump stars (RC) from those that are still on their first ascent of the red giant branch (RGB). We extend these studies to a tenfold larger wavelength range of $0.33$ to $2.5\mu$m with the moderate-resolution VLT/X-shooter spectrograph. Our analysis of 49 stars with asteroseismic data from the \ktwo\ mission confirms that CN, CO and CH features are indeed the primary carriers of spectroscopic information on the evolutionary stages of red giant stars. We report 215 informative features for differentiating the RC from the RGB within the range of $0.33$ to $2.5\mu$m. This makes it possible for existing and future spectroscopic surveys to optimize their wavelength regions to deliver both a large variety of elemental abundances and reliable age estimates of luminous red giant stars.
\end{abstract}


%%%%%%%%%%%%%%%%%%%%%%%%%%%%%%%%%%%%%%%%%%%%%%%%%%

%%%%%%%%%%%%%%%%% BODY OF PAPER %%%%%%%%%%%%%%%%%%

\section{Introduction}

The physical processes that drive the assembly and evolution of spiral galaxies leave imprints on the distributions of stellar age, abundance, and kinematics.
Mapping those observable properties in spiral galaxies aids in inferring their formation histories \citep[e.g.,][]{Poci19} and how they have evolved based on factors such as mass and environment \citep[e.g.,][]{Santucci20}.
Observations of stars in the Milky Way provide complementary information, allowing a highly detailed look into distinct nucleosynthetic sites and processes for dynamical evolution \citep[e.g.,][]{Lucey22,Chen22}.

The combination of large-scale spectroscopic surveys and the \gaia\ astrometric mission \citep{Gaia16, Gaia22} has revolutionised this field within the Milky Way, making it possible to map chemo-dynamical structure in great detail \citep[e.g.,][]{Casey17, Grieves18,Hayden18, Buder19, BlandHawthorn19, Hayden20,Medan23}. However, this 
detailed view is restricted to a zone of $\sim3$~kpc around the Sun, where the measurement uncertainty in \gaia\ parallax produces a fractional uncertainty in distance of 10\% or less (e.g. \citealt{Luri18}). Using stellar standard candles as a tracer population is one way to extend precise chemodynamic investigations to more distant regions of the Galaxy. The red clump (RC), which is a post-red giant branch (RGB) phase of core helium burning, is one point in stellar evolution that can be used as a standard candle \citep[e.g.,][]{Cannon70, Girardi16}. 
This is due to the fact that all low-mass ($0.8-2.0\rm{M}_\odot$), non-metal-poor ([Fe/H] $>-0.5$) RGB stars ignite helium fusion when their core reaches a threshold mass of 0.47 $M_\odot$, independent of the original stellar mass \citep{Girardi16}.
As a result, these stars all transition into RC stars with identical helium-burning core masses and any luminosity difference between them, which is small, arises from differences in their envelope masses and compositions. 

% \begin{table*}
%     \centering
%     \caption{Gaia DR3 and EPIC ID numbers, on-sky coordinates, observation date, and exposure time for our program targets. The full table is available in the online version of this article; an abbreviated version is included here to demonstrate its form and content.}
%      \label{tab:theobs}
%         \begin{tabular}{l|l|c|c|l|l|l|l}
%         \hline
%         \gaia\ DR3 ID & EPIC ID & $\alpha (^{\circ})$ & $\delta (^{\circ}) $ & UT date & t$_{\rm exp,UBV}$(s) & t$_{\rm exp,VIS}$(s) & t$_{\rm exp,NIR}$(s) \\
%         \hline
%         3601038161655293056 & 201244747 & 179.0137 & -3.2889 & 11-Feb-2019 & 260 & 240 & 80 \\
%         3601940791982219008 & 201300337 & 179.8983 & -2.4452 & 11-Feb-2019 & 300 & 280 & 100 \\
%         3793639613492087040 & 201349077 & 173.1241 & -1.7199 & 08-Jan-2019 & 260 & 260 & 80 \\
%         \hline        
%         \end{tabular}
% \end{table*}

% \begin{table}
%     \centering
%     \caption{EPIC ID numbers, \teff, \logg, [Fe/H], \dnu, and  \texttt{RC\_Prob} for our program targets. The full table is available in the online version of this article; an abbreviated version is included here to demonstrate its form and content.}
%     \label{tab:parameters}
%     \begin{tabular}{lrrrrr}
%     \hline
%        EPIC-ID &      \teff\ (K) &  \logg &      [Fe/H] & \dnu\ ($\mu$Hz) &  \texttt{RC\_Prob} \\
%        \hline
% 201208734 & 4714 &  2.375 &  0.475 & 4.76 &    0.875 \\
% 201244747 & 4662 &  2.220 & -0.415 & 5.63 &    0.000 \\
% 201245001 & 4712 &  2.895 &  0.467 & 5.78 &    0.000 \\
% \hline
% \end{tabular}
% \end{table}

\begin{table*}
    \centering
    \caption{EPIC ID numbers, on-sky coordinates, observation date, exposure time, stellar parameters (i.e. \teff, \logg, [Fe/H]) and asteroseismic labels (i.e. \dnu\ and \texttt{RC\_Prob}) for our program targets. The full table is available in the online version of this article; an abbreviated version is included here to demonstrate its form and content.}
     \label{tab:theobs}
        \begin{tabular}{l|c|c|l|l|l|l|l|l|l|l|l}
        \hline
        EPIC ID & $\alpha (^{\circ})$ & $\delta (^{\circ}) $ & UT date & t$_{\rm exp,UBV}$(s) & t$_{\rm exp,VIS}$(s) & t$_{\rm exp,NIR}$(s) & \teff\ (K) & \logg & [Fe/H] & \dnu\ ($\mu$Hz) & \texttt{RC\_Prob}\\
        \hline
        201244747 & 179.0137 & -3.2889 & 11-Feb-2019 & 260 & 240 & 80 & 4662 & 2.220 & -0.415 & 5.63 & 0.000 \\
        201300337 & 179.8983 & -2.4452 & 11-Feb-2019 & 300 & 280 & 100 & 4816 & 2.445 & -0.455 & 4.93 & 0.245 \\
        201349077 & 173.1241 & -1.7199 & 08-Jan-2019 & 260 & 260 & 80 & 4668 & 2.651 & -0.173 & 5.79 & 0.038 \\
        \hline        
        \end{tabular}
\end{table*}

A challenge when using RC stars as standard candles is that they occupy a similar colour and magnitude space as non-standard candle stars on the lower RGB. While RC stars are well confined in luminosity, with $\log(L_{\rm{ RC}}/L_{\odot})=1.95$ and a mean intrinsic dispersion of $\sim0.17\pm0.03$\,mag in all colour bands \citep{Hawkins17}, RGB stars with similar colours and effective temperatures occupy a wider range in luminosity. Even with spectroscopic measurements of \logg\ confining the luminosity estimates of RGB stars, their distance estimates have an uncertainty of up to $\sim 10$\% \citep[e.g.][]{Bovy14}.

Methods for probabilistically separating red clump and red giant branch stars on the basis of spectroscopic stellar parameters have been developed by a number of authors, including \citet{Williams13}, \citet{Bovy14} and \citet{Sharma18}. Many authors, for example \cite{Wan15}, find that these methods can result in up to 20\% contamination, i.e. RGB stars being misidentified as RC stars. Fortunately, asteroseismology can unambiguously distinguish the two types of stars \citep{Bedding11} using the frequency separation of overtone acoustic oscillations \dnu\ and the period spacing between gravity oscillations $\Delta$P. However, accurate asteroseismic analysis of the oscillations present in red giant stars requires a well-sampled series of precise measurements of either radial velocity or photometry across a sufficiently long time baseline \citep{Hon18}. 

Thanks to the \kepler, \ktwo, and \tess\ space missions \citep[respectively]{Borucki10,Howell14,Ricker15} the availability of asteroseismic data across the sky has increased dramatically, and methods for quickly and reliably extracting asteroseismic parameters from those data are emerging (e.g., \citealt{Hon22,Zinn22,Reyes22}). However, spectroscopic data sets contain more stars spanning a larger volume in the Galaxy, and they will continue to do so into the future \citep{Martell21}.

Two recent papers demonstrated that the asteroseismic parameters \dnu\ and $\Delta$P can be inferred from stellar spectroscopy, using data from the APOGEE ($R \approx 20,000$ covering $1.5-1.7 \mu$; \citealt{Hawkins18}) and LAMOST ($R \approx 1800$ with full optical coverage; \citealt{Ting18}) surveys. Both studies used samples of stars with clear asteroseismic identifications from \ktwo\ along with spectroscopy, with \citet{Hawkins18} training the data-driven regression tool \cannon\ \citep{Ness15} to predict \dnu\ and $\Delta$P and \citet{Ting19} training a neural network for the same purpose. Both methods produced RC/RGB classifications with $\approx 3\%$ error rates, which is distinctly better than the $10 - 25\%$ misclassification rates for methods based on non-seismic measurements \citep[see the discussion on RGB contamination in][]{Bovy14}.
\citet{Hawkins18} found that the model produced by \cannon\ showed sensitivity to CN and CH molecular absorption features, suggesting that 
deep mixing beyond the first dredge-up
as red giants ascend the RGB \citep[e.g.,][]{Martell08, Placco14, Shetrone19} or a sharp change to surface abundances at the helium flash might be responsible. However, with the limited spectral coverage of APOGEE data, they were not able to investigate the origins of the spectroscopic differences more thoroughly.

To obtain a broader view of which features in the spectrum differentiate RC and RGB stars we carried out an observing programme with the X-shooter spectrograph \citep{Vernet11}, which is mounted on UT3 (Melipal\footnote{This is the traditional name of the Southern Cross in Mapuche language, the Indigenous peoples of south-central Chile and southwestern Argentina.}) at the ESO Very Large Telescope in Chile. 
We collected $R\sim10,000$ spectra with full wavelength coverage from $0.33$ to $2.5\mu$m for 49 stars in the overlapping region of the RC and RGB in temperature-luminosity space with asteroseismic classifications from \ktwo.
In this Letter we analyse these spectra to determine the significant spectral differences between RC and RGB stars and identify the spectral features that are most effective for their classification without the need for asteroseismic data.

\section{The data set}
\label{sec:data}
The target list for this study was assembled in early 2018 using the asteroseismic parameters available from the \ktwo\ mission \citep{Stello17} at the time. We selected 50 stars near the red clump in stellar parameter space across \ktwo\ Campaigns 1, 4 and 5. The stars were selected with 
$4200 \leq T_{\rm eff} \leq 5200$, $1.8 \leq \text{log} g \leq 3.2$, and $-0.45 \leq \rm{[Fe/H]} \leq +0.5$. 
Half the stars are classified as RC stars and half are RGB stars by the method detailed in \citet{Hon18} using the input of \dnu\ from the SYD pipeline \citep{Huber09}. 
Utilising the mass scaling relations identified in \cite{Zinn22}, we find that these stars have masses in the range $0.9\leq M/M_\odot\leq 1.9$. 

Of these 50 initial stars, 49 received X-shooter observations in service mode under sufficiently good conditions for science use in our program 0102.D-0517. All stars were observed with a narrow slit ($0\farcs 4$, $0\farcs 5$, and $0\farcs 5$ for the UV, visible, and IR arms of the spectrograph, respectively), with nodding along the slit to facilitate sky subtraction. Data were read out in 100\,kHz mode with 1x1 binning, and exposure times were chosen to achieve a signal-to-noise ratio per pixel of 150 in each arm. Observing details are listed in Table \ref{tab:theobs}.

Analysis of the \ktwo\ data has improved since our initial target selection in 2018, hence we use the latest \ktwo\ asteroseismic parameters from \cite{Zinn22} in this work, incorporating the large frequency separation \dnu\ determined with the SYD pipeline \citep{Huber09}. We verify the reliability of \dnu\ with the \texttt{dnu\_prob} score from \cite{Reyes22}, estimated using a neural network classifier. This ensures that a given \dnu\ value derived from time-series observations is reliable, regardless of the method used for its estimation. In this data set 32 stars have $\texttt{dnu\_prob}>0.6$ and hence their \dnu\ values and RC/RGB classifications are reliable. We also use the RC and RGB classifications determined from the neural network outlined in \cite{Hon18} (i.e. \texttt{RC\_Prob}), which holds information about the period spacing, $\Delta$P, the key seismic parameter for RC-RGB classification. The \texttt{RC\_Prob} value is a probability that a star is in the RC (i.e. $\texttt{RC\_Prob}\geq0.5$) or the RGB (i.e. $\texttt{RC\_Prob}<0.5$) phase. The closer a star lies to the extremes of this probability, the more confident that classification is. Table \ref{tab:theobs} also lists ID numbers, \teff, \logg, [Fe/H], \dnu, and \texttt{RC\_Prob} for the program stars.

\subsection{Spectroscopic Data Reduction}

Data reduction for the X-shooter spectra was performed with the X-shooter pipeline version 3.2.0 \citep{Modigliani10} using the ESO Reflex environment \citep{Freudling13}, developed by the European Southern Observatory for reducing VLT/VLTI science data. Using the standard settings, the pipeline processes data from each spectrograph arm separately, proceeding through bias subtraction, flat-field correction, wavelength calibration based on arc lamp exposures, extraction of the 1D spectrum, and flux calibration using a standard star. 
We corrected the spectra for telluric absorption using molecfit \citep{Kausch15,Smette15} in a similar approach as detailed in \cite{Gonneau20}. First, we apply molecfit to the entire spectrum to derive the precipitable water vapor column (PWV). Then we divide the spectrum into smaller wavelength segments and apply molecfit locally using the determined PWV value, 
in order to find better local wavelength solutions and line spread functions. 
The corrected wavelength segments are then merged together.

Regions of the spectra with strong telluric features, and regions where the transmission efficiency of the dichroic varies strongly, were masked out before continuum normalisation, which was carried out for each spectrograph arm separately. This masking mainly affected the NIR arm and the extremes of the three arms \citep{Verro21}.
Following continuum normalisation, the three spectrograph arms were combined into a single spectrum for each star and interpolated onto a single wavelength array to permit analysis with \cannon.

\begin{figure}
    \centering
    \includegraphics[width=\columnwidth]{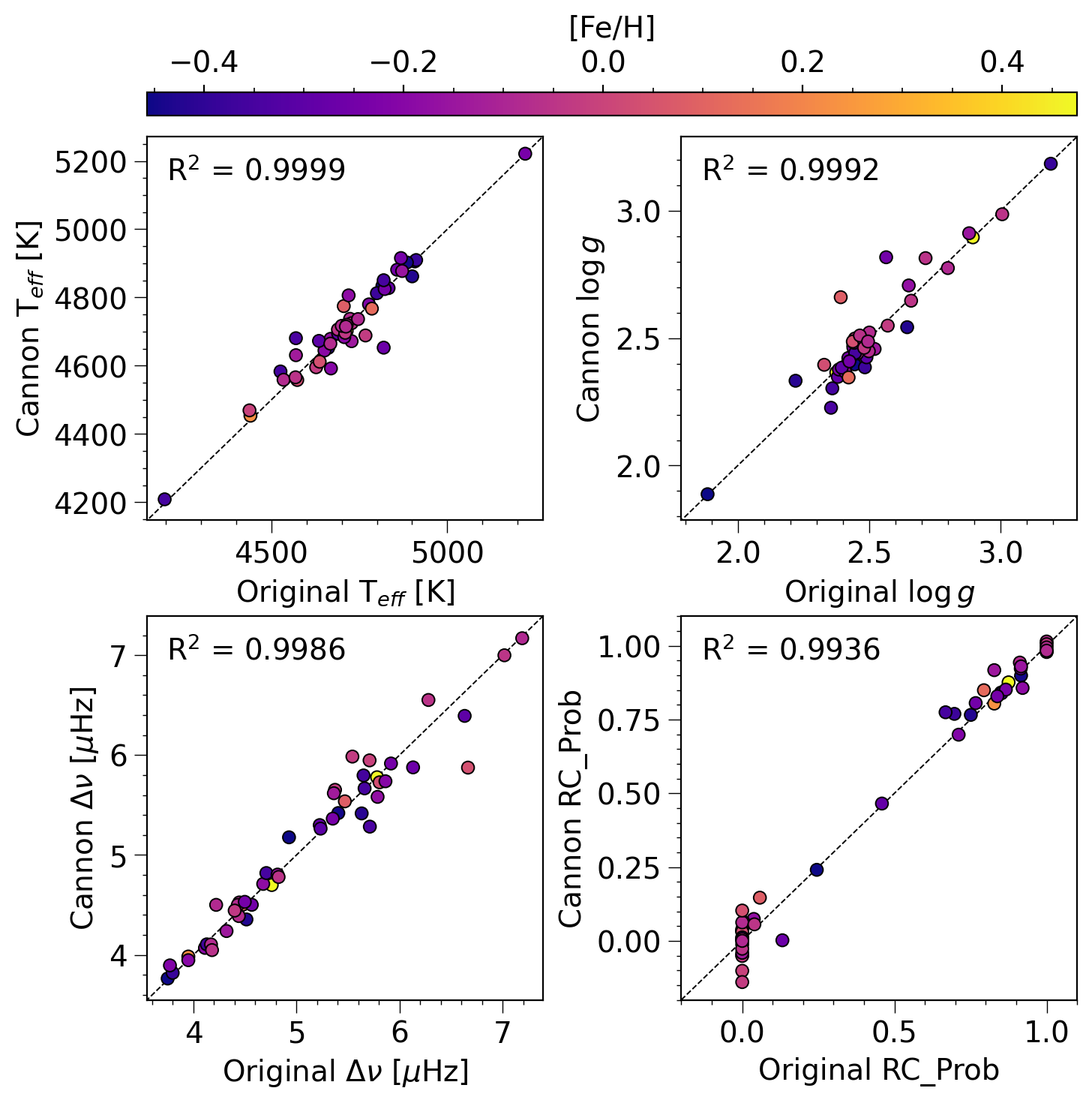}
    \caption{In this figure we show the capability of \cannon\ to predict the labels of the model. The x-axis of each plot illustrates the known values of each parameter and the y-axis shows the values predicted by \cannon. The dashed diagonal lines show the one-to-one relationship between the known and predicted values. Here we also include the $R^2$ value of the one-to-one fit with the data. Even with a small data set of 49 stars our model is able to predict the parameters we train the model on with decent accuracy.}
    \label{fig:one-to-one}
\end{figure}

\section{Analysis with The Cannon}

\cannon \footnote{The \cannon\ is named after Annie Jump Cannon, an inspirational woman in astronomy who was able to classify stars at impressive speed, not from physical models but from their spectra alone.} \citep{Ness15,Casey16} is a computationally efficient data-driven algorithm that generates a model from observed stellar spectra as a quadratic function of user-defined input parameters called \textit{labels}. \cannon\ works in two main steps: \textit{training} and \textit{test}. In the \textit{training step}, a model is generated from a selection of stars with known properties, called \textit{training objects}. This model 
consists of the coefficients of the quadratic terms that best reproduce the the flux at each wavelength pixel for the training objects. The original paper on \cannon\ by \citet{Ness15} demonstrates that the stellar parameters \teff, \logg, and [Fe/H] are effective labels for many normal stars. In the \textit{test step}, labels are inferred for the \textit{test objects} using the trained model. 

Our use of \cannon\ involves generating a model that uses both stellar and asteroseismic parameters as labels, i.e. \teff, \logg, [Fe/H], \dnu\ and \texttt{RC\_Prob}, to investigate the sensitivity of red giant spectra to the classification of RC and RGB stars. 
The asteroseismic parameter \texttt{RC\_Prob} is the main focus of this study, however, we also include \dnu\ and the stellar parameters \teff, \logg, and [Fe/H] from GALAH data release 3 \citep{Buder21} because they also influence the flux, and the data set spans a small but nonzero range in each of these dimensions.

While 32 stars in the data set have $\texttt{dnu\_prob}>0.6$ according to \citet{Reyes22}, i.e. their \dnu\ values are reliable, the remaining 17 stars and their respective \dnu\ values are not. We  investigate these stars by training \cannon\ with the 32 reliable stars and performing the \textit{test step} on the remaining 17 stars. We find that \cannon\ predicts \dnu\ values for these stars that are quite similar to the catalogue values (mean offset 0.1, standard deviation 0.42), and so we infer that those catalogue values (from \citealt{Zinn22}) are also reliable for the purpose of this work. Hence we train \cannon\ with all 49 stars of the data set.

\begin{figure*}
    \centering
    \includegraphics[width=\textwidth]{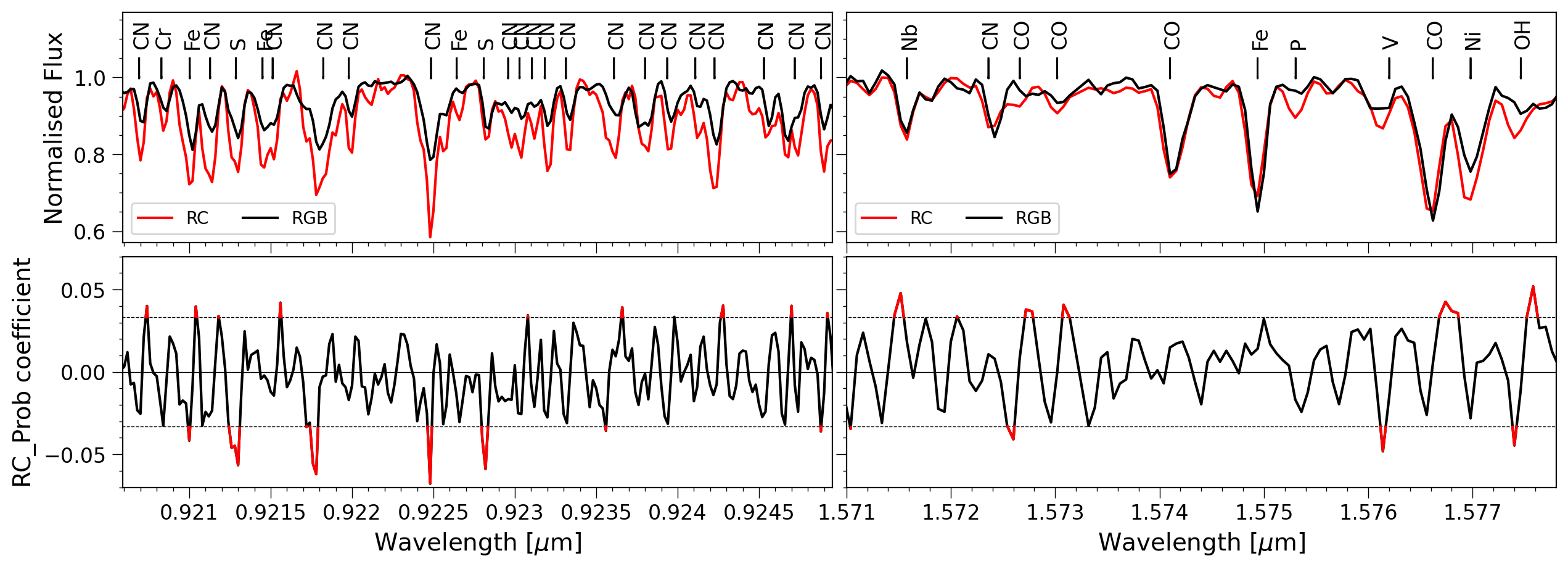}
    \caption{In this Figure we show a comparison between the spectra of one RC star (EPIC-211413402; red line in the top panel; $T_{\rm{eff}}=4668,\log g=2.5,\rm{[Fe/H]}=-0.08,\Delta\nu=4.44,\rm{\texttt{RC\_Prob}}=1.0$) and one RGB star (EPIC-201244747; black line in the top panel; $T_{\rm{eff}}=4662,\log g=2.2,\rm{[Fe/H]}=-0.4,\Delta\nu=5.6,\rm{\texttt{RC\_Prob}}=0.0$) across two different spectral windows to demonstrate the significance of particular CN and CO features to the classification of red giant stars. The bottom panels illustrate the significance of each wavelength pixel through the linear term coefficient of \cannon\ model pertaining to \texttt{RC\_Prob}. Pixels with values around zero (the solid black line) hold little to no significance whereas pixels with values above 0.033 or below -0.033 (highlighted in red) indicate a significance of greater than the 90th percentile. The left panels illustrate various CN features that \cannon\ model found to be effective in red giant classification and the right panels illustrate a few CO as well as CN and OH molecular features.}
    \label{fig:ex_lines}
\end{figure*}

%############### POTENTIALLY CUT THIS SECTION FROM HERE ###############

\cannon\ is often used to infer stellar labels. While this isn't the focus of this investigation we demonstrate this use on our data set in Figure \ref{fig:one-to-one}. Here we show the parameters predicted by \cannon\, i.e., \teff, \logg, \dnu\ and \texttt{RC\_Prob}, for the whole data set after being trained also on the whole data set. We note that this isn't common practice, however, we validate the robustness of our model by performing a series of ``leave-one-out'' tests, such that we additionally set up and trained \cannon\ with all but one star in our data set. We repeated this 49 times, removing a different star on each iteration, thus resulting in 49-fold cross-validation. We find that the 90th percentile of the distribution of the \texttt{RC\_Prob} linear coefficient does increase slightly from $\pm0.033$ in the model generated with the full data set due to there being fewer stars in the training set. However, we note that regardless of which star is removed from the \textit{training objects}, the 90th percentile of the distribution of \texttt{RC\_Prob} coefficients does not change significantly between models with a mean of $0.0428${\raisebox{0.5ex}{\tiny$^{+0.006}_{-0.007}$}}.

In each panel of Figure \ref{fig:one-to-one}, the x-axis represents the known values for each parameter and the y-axis represents the values predicted by \cannon. The dotted lines on each panel show a one-to-one relationship between the known and predicted values. We performed a linear regression on the predicted values and present the $R^2$ values of the one-to-one fit for each parameter in each panel of Fig. \ref{fig:one-to-one}. Even with a small data set of 49 stars, \cannon\ is capable of accurately predicting stellar parameters without the need for synthetic spectral models.

%############### POTENTIALLY CUT THIS SECTION TO HERE ###############

In this investigation, we use \cannon\ to identify which spectral features are most informative for classifying RC and RGB stars. We do this by investigating the coefficients of the quadratic function that predicts the flux value for each pixel. Our model contains 20 terms; five linear terms, ten cross terms and five quadratic terms. The linear term coefficients indicate the lowest-order dependence of the flux for each label, i.e. they indicate the weight each wavelength pixel has for predicting each label \citep{Casey16}.
These terms are a good first indicator of the sensitivity of a pixel in predicting a particular label. In the particular case of classifying RC and RGB stars, we use the linear term \texttt{RC\_Prob} for this purpose.

\begin{table}
\centering
\caption{A representative list of the significant features identified by \cannon\ for the classification of red giants including their location in the wavelength range, the value of the coefficient pertaining to the linear \texttt{RC\_Prob} term as well as the percentile of these values. A full table detailing all of the significant features identified in this work is available in the online version of this article.}
\label{tab:top10}
\begin{tabular}{lrrr}
\hline
Feature &  Wavelength ($\mu$m) & \texttt{RC\_Prob} Coeff &  Percentile (\%) \\
\hline
CN Violet 3-3 &  0.386 &          0.128 &       99.53 \\
   CN Red 5-1 &  0.668 &         -0.126 &       99.07 \\
         Ni I &  0.762 &         -0.108 &       98.60 \\
$^{13}$CO 3-2 &  2.063 &         -0.106 &       98.14 \\
$^{13}$CO 3-2 &  2.070 &         -0.105 &       97.67 \\
         Ti I &  0.399 &          0.100 &       97.21 \\
CN Violet 0-0 &  0.377 &         -0.097 &       96.74 \\
   CH B-X 0-0 &  0.391 &         -0.089 &       96.28 \\
   CN Red 4-0 &  0.626 &          0.087 &       95.81 \\
       CO 8-7 &  2.067 &         -0.085 &       95.35 \\
\hline
\end{tabular}
\end{table}

\section{Spectroscopic features of interest}
We find a number of spectral features to be particularly relevant for distinguishing RC stars from the RGB. A spectral feature is considered significant if the linear coefficient \texttt{RC\_Prob} pertaining to that feature is above the 90th percentile of the distribution, which is $\pm 0.33$. 
We also filter for spurious instances that are not indicative of line strength variation, for example, when the value of the \texttt{RC\_Prob} coefficient exhibits a narrow, strongly negative feature immediately followed by a narrow, strongly positive feature or vice versa.

There are both atomic and molecular features that we find as significant in the classification of red giant stars 
from the linear term coefficient \texttt{RC\_Prob}. In the region of the spectrum immediately around these features that we find to be significant for RC-RGB classification, the linear and square terms for the \texttt{RC\_Prob} label (i.e. \texttt{RC\_Prob} and \texttt{RC\_Prob}$^2$) are highly correlated, thus validating their identification as significant.
The molecular features identified in our model include CN, CO and CH. This includes a total of 64 CN, 41 CO and 27 CH features. The CH features are found predominantly at the blue end of the spectrum and the CO features dominate in the infrared beyond $1.5\mu$m, while the CN features are distributed broadly throughout the wavelength range.

Figure \ref{fig:ex_lines} illustrates some of these features over two spectral windows ($0.920\,\mu\rm{m}-0.925\,\mu\rm{m}$ in the left panels, and $1.571\,\mu\rm{m}-1.578\,\mu\rm{m}$ in the right panels). 
The top panels compare the spectra of two stars in our data set with the most similar parameters, they are the RC star EPIC-211413402 with $T_{\rm{eff}}=4668,\log g=2.5,\rm{[Fe/H]}=-0.08,\Delta\nu=4.44,\rm{\texttt{RC\_Prob}}=1.0$ (red line) and the RGB star EPIC-201244747 with $T_{\rm{eff}}=4662,\log g=2.2,\rm{[Fe/H]}=-0.4,\Delta\nu=5.6,\rm{\texttt{RC\_Prob}}=0.0$ (black line).
We also identify both atomic and molecular features with arrows and their appropriate labels. The features in the left panel were sourced from the open-source line list generator \texttt{linemake} \citep{Placco21}, while those in the right panel were derived from the APOGEE DR16 spectral line list \citep{Smith21}. The bottom two panels show the linear term coefficient of \cannon\ model pertaining to the label \texttt{RC\_Prob}, illustrating the sensitivity of each wavelength pixel to the prediction of \texttt{RC\_Prob}.

The left panels highlight numerous CN features that have been identified by \cannon\ model as significant in the classification of RC from RGB stars. There are compelling differences in the spectra of the RC star compared to the RGB star, particularly in the CN features at $\sim0.9218\mu$m and $\sim0.9225\mu$m, despite possessing similar \teff, \logg, and [Fe/H]. The right panels highlight a few CO features, as well as another CN feature and an OH feature, that can be used to effectively classify RC stars from RGB stars. While the differences between the spectra of the RC and RGB stars in this spectral window are more subtle, they are still useful to a trained \cannon\ model in predicting the stellar classification of red giants.

Numerous atomic features are also identified as significant for classifying RC and RGB stars. The vast majority of these atomic features are iron with a total of 38 features spread throughout the optical and near-infrared parts of the spectrum ($<0.9\mu$m). Some less numerous, but still significant, atomic features include titanium, and the iron peak elements vanadium and nickel. 
However, we note that many of these atomic features have significant covariance with other coefficients such as \logg\ and \teff\ in particular. Therefore, these atomic features may not be as useful as \cannon\ implies in the classification of red giants. We do not find significant covariance with the CN and CO features identified in this model.

% However, we note that many of these features also hold high significance in predicting \teff\ and \logg, therefore they may not be as useful as \cannon\ implies in the classification of red giants.

Table \ref{tab:top10} presents 
a representative list of the significant features in order of significance that \cannon\ identified for differentiating RC and RGB stars, including their positions in the wavelength range, the value of their \texttt{RC\_Prob} coefficient, and the percentile rank of these coefficient values, indicating their overall importance. The identified features are mainly CN and CO molecular features. A full table detailing all of the significant features identified in this work is available in the online version of this article.

\section{Discussion and future work}
Here we demonstrate that CN, CO and CH molecular features hold the most significance in the classification of red giant stars based solely on their spectra. This work builds upon previous spectro-seismic investigations, expanding the wavelength range to cover $0.33$ to $2.5\mu$m with moderate-resolution VLT/X-Shooter spectra. We used the data-driven algorithm \cannon\ to generate a quadratic model from the observed stellar spectra of 49 red giant stars taking into account their stellar parameters \teff, \logg, and [Fe/H] \citep[sourced from the GALAH survey;][]{Buder21} and asteroseismic labels \dnu\ and \texttt{RC\_Prob} \citep[sourced from the \ktwo\ mission;][]{Zinn22}.

Many significant features identified in this study also fall within the wavelength ranges of well-established spectroscopic surveys with higher resolution, including GALAH \citep{Buder21} and APOGEE \citep{Majewski17}. From the total of 215 significant features identified in this study, 23 are within the wavelength range of the GALAH survey, 14 features (mostly CO and CN) are within the wavelength range of the APOGEE survey and 105 features (dominated by CN) are within the wavelength range covered by the Veloce spectrograph \citep[$580-930$ nm;][] {Gilbert18}.

This work supports the results of \cite{Hawkins18} and further demonstrates that optical and infrared spectra of red giant stars are imprinted with asteroseismic information which can be used to classify red giant stars.
This allows for the reliable selection of RC stars from existing large spectroscopic surveys, such as GALAH and APOGEE, for use in Galactic archaeology investigations probing the history and structure of the Milky Way galaxy.

To understand the physical origins of this spectro-seismic relation, we will need to look at models of RGB evolution and the helium flash. The presence of $^{13}$CO in these significant features is suggestive of deep mixing as a key driver of the spectroscopic differences between red giant stars, and offers clear avenues for further investigation. We are currently undertaking a higher-resolution study of this problem with a larger sample. This will provide us with clearer insight into the evolution of CN abundance between the RC and RGB stellar phases. 

In addition, this will hopefully provide additional insights into the physical process at work during this evolution. This deeper study investigates stars selected from GALAH \citep{Buder21} with \ktwo\ and \tess\ asteroseismology \citep[][respectively]{Howell14,Ricker15} and higher resolution spectra obtained with the Veloce spectrograph ($R\sim80,000$) at the Anglo-Australian Telescope \citep{Gilbert18}. This higher-resolution investigation will also allow us to expand the suite of elemental abundance information for these stars and derive asteroseismic ages in addition to the information that is already available from GALAH.

\section*{Acknowledgements}

KAB, CYYH and SLM acknowledge funding support from the UNSW Scientia program. SLM is supported by the Australian Research Council through Discovery Project grant DP180101791. This work was supported by the Australian Research Council Centre of Excellence for All Sky Astrophysics in 3 Dimensions (ASTRO 3D), through project number CE170100013.

This work is based on observations collected at the European Organisation for Astronomical Research in the Southern Hemisphere under ESO programme 0102.D-0517(A). This paper includes data collected by the Kepler mission and obtained from the MAST data archive at the Space Telescope Science Institute (STScI). Funding for the Kepler mission is provided by the NASA Science Mission Directorate. STScI is operated by the Association of Universities for Research in Astronomy, Inc., under NASA contract NAS 5–26555.

The authors acknowledge the Traditional Custodians of the land on which the analysis for this investigation was conducted, the Bedegal people of the Eora nation. The authors pay respect to elders both past and present and extend that respect to other Aboriginal and Torres Strait Islander peoples reading this paper.

The authors also thank the anonymous reviewer for their helpful feedback which played an important role in improving this manuscript. 

%%%%%%%%%%%%%%%%%%%%%%%%%%%%%%%%%%%%%%%%%%%%%%%%%%
\section*{Data Availability}

The data discussed in this Letter were acquired through ESO programme 0102.D-0517(A) and are available in the ESO archive.
%The inclusion of a Data Availability Statement is a requirement for articles published in MNRAS. Data Availability Statements provide a standardised format for readers to understand the availability of data underlying the research results described in the article. The statement may refer to original data generated in the course of the study or to third-party data analysed in the article. The statement should describe and provide means of access, where possible, by linking to the data or providing the required accession numbers for the relevant databases or DOIs.



\bibliographystyle{mnras}
\bibliography{1_citations} % if your bibtex file is called example.bib

\bsp	% typesetting comment
\label{lastpage}
\end{document}